\documentclass{article}

\usepackage{microtype}
\usepackage{graphicx}
\usepackage{subfigure}
\usepackage{booktabs} 

\usepackage{hyperref}

\usepackage{enumitem}

\usepackage[scaled=0.86]{helvet}

\usepackage[table]{xcolor}
\usepackage[accepted]{icml2024}

\usepackage{amsmath}

\usepackage{amssymb}
\usepackage{mathtools}
\usepackage{amsthm}
\usepackage{xspace}
\usepackage{multirow}
\usepackage{pgf}

\usepackage[capitalize,noabbrev]{cleveref}

\usepackage{listings}
\usepackage[scaled=0.95]{inconsolata}

\theoremstyle{plain}

\theoremstyle{definition}

\theoremstyle{remark}

\icmltitlerunning{Do Large Code Models Understand Programming Concepts? Counterfactual Analysis for Programming Predicates}

\begin{document}

\twocolumn[
\icmltitle{Do Large Code Models Understand Programming Concepts?\\ Counterfactual Analysis for Code Predicates}



\icmlsetsymbol{equal}{*}

\begin{icmlauthorlist}
\icmlauthor{Ashish Hooda}{uwm,google}
\icmlauthor{Mihai Christodorescu}{google}
\icmlauthor{Miltiadis Allamanis}{dm}
\icmlauthor{Aaron Wilson}{google}
\icmlauthor{Kassem Fawaz}{uwm}
\icmlauthor{Somesh Jha}{uwm,google}
\end{icmlauthorlist}

\icmlaffiliation{uwm}{UW-Madison}
\icmlaffiliation{google}{Google Research}
\icmlaffiliation{dm}{Google DeepMind. \\The majority of this work
was completed while Ashish Hooda was an intern at Google Research.}


\icmlcorrespondingauthor{Ashish Hooda}{ahooda@wisc.edu}

\icmlkeywords{Machine Learning, ICML}

\vskip 0.3in
]



\printAffiliationsAndNotice{}  

\newcommand{\name}{\lstinline{CACP}\xspace}
\newcommand{\ah}[1]{\todo[AH]{#1}}
\newcommand{\mc}[1]{\todo[Mihai]{#1}}

\definecolor{BaseRed}{rgb}{1,0,0}
\definecolor{BaseGreen}{rgb}{0,1,0}

\newcommand{\scalefactor}{50} 

\newcommand{\intensitycolor}[3]{
    \ifdim #1 pt > 0 pt
        \expandafter\cellcolor\expandafter{BaseRed!#2!white}#2 \%
    \else
        \expandafter\cellcolor\expandafter{BaseRed!#2!white}#2 \%
    \fi
}

\newcommand{\ratiocolor}[2]{
    \ifdim #1 pt > 0 pt
        \expandafter\cellcolor\expandafter{BaseRed!#2!white}#1 
    \else
        \expandafter\cellcolor\expandafter{BaseGreen!#2!white}#1 
    \fi
}



\newcommand{\predicate}{\lstinline{PCP}\xspace}

\newcommand{\predicates}{\lstinline{PCPs}\xspace}
\begin{abstract}
    Large Language Models' success in text generation has also made them better at code generation and coding tasks. While a lot of work has demonstrated their remarkable performance on tasks such as code completion and editing, it is still unclear as to why. We help bridge this gap by exploring to what degree auto-regressive models understand the logical constructs of the underlying programs. We propose Counterfactual Analysis for Programming Concept Predicates (\name) as a counterfactual testing framework to evaluate whether Large Code Models understand programming concepts. With only black-box access to the model, we use \name to evaluate ten popular Large Code Models for four different programming concepts. Our findings suggest that current models lack understanding of concepts such as data flow and control flow. 
\end{abstract}
\section{Introduction}

Language Language Models (LLMs) have demonstrated remarkable performance on a variety of automated programming tasks, such as code completion~\cite{austin2021program,fried2022incoder}, code repair~\cite{jiang2021cure,joshi2023repair}, and code translation~\cite{pan2023understanding,chen2023teaching}. Automating a programming task is a complex problem that requires understanding many concepts in the underlying code. These concepts include how variables are stored, accessed, and modified in memory; how execution proceeds across various constructs; and how different parts of the code compose sequentially or in parallel to perform a computation. We refer to these concepts as \textit{Programming-Concept Predicates} (\predicates). Despite their remarkable performance, the degree to which LLMs understand the \predicates in the programs they manipulate remains unclear.

Empirical evaluations on benchmark datasets such as HumanEval~\cite{chen2021evaluating}, MBPP~\cite{austin2021program}, and CodeContests~\cite{doi:10.1126/science.abq1158} drive the current understanding of the code capabilities of LLMs. While task-driven evaluation measures the end-to-end performance, it does not reveal the LLM's fine-grained understanding of \predicates. As a result, we often cannot attribute the failures in these coding tasks to specific aspects of the underlying code --- was the code completion wrong due to confusing variable names, unusual control flow, inherent algorithmic complexity, or code size? Such a fine-grained attribution would allow practitioners to better reason about these models' limits and highlight the avenues to improve their performance.

In this work, we consider the problem of \textit{evaluating a given model's understanding of programming concepts}. We focus on four \predicates that represent classical concepts in the program analysis literature~\cite{allen1970control,fosdick1976data,lin2008evaluation,dart1992efficient}:
\begin{itemize}[itemsep=0.2ex,topsep=0.25ex,leftmargin=1.5em]
    \item[$\qed$] \textit{Control Flow:} The output of the automated coding task does not change with the ordering of independent code statements.
    \item[$\qed$] \textit{Data Flow:} The automated coding task uses only variables that are in scope (and live) within the coding task.
    \item[$\qed$] \textit{Data Types:} The automated coding task satisfies the constraints of the type system.
    \item[$\qed$] \textit{Identifier Naming:} Functionality of the automated coding task does not depend on the names of the variables or functions.
\end{itemize}

We introduce Counterfactual Analysis for Programming Concept Predicates (\name), a counterfactual analysis framework for evaluating whether large code models understand \predicates. As the name suggests, \name builds on counterfactual analysis to cast concept understanding as the problem of determining how controlled input changes result in model output changes. There are two main components of \name -- (1) Generating counterfactuals for code that only perturb specific \predicates, and (2) Using them to analyze the model's performance. Specifically for a given \predicate, we define code perturbations (called \textit{mutations}) that are minimal in that they influence only one \predicate, but not others. The challenge lies in defining these minimal mutations and predictably evaluating their impact on the model output. The minimality of mutations allows us to explain failures concerning specific \predicates that are not well understood by the model.

We apply our \name framework on code completion (the most popular code task for language models) and show how to benchmark predicate understanding with only hard-label black-box access to a model. This allows us to quantify the model's coding capability through an end-to-end automated measurement of understanding of \predicates related to the task, without having to adapt the model to those predicates (e.g., without fine-tuning or using additional training data). We develop four mutations that instantiate the \predicates described above: flipping if-else conditions, swapping independent statements, breaking def-use chains, and changing variable names. Building on these mutations, we create a new benchmark dataset to evaluate how LLMs understand \predicates. 

\begin{figure}[t]
  \centering

  \includegraphics[width=0.75\columnwidth]{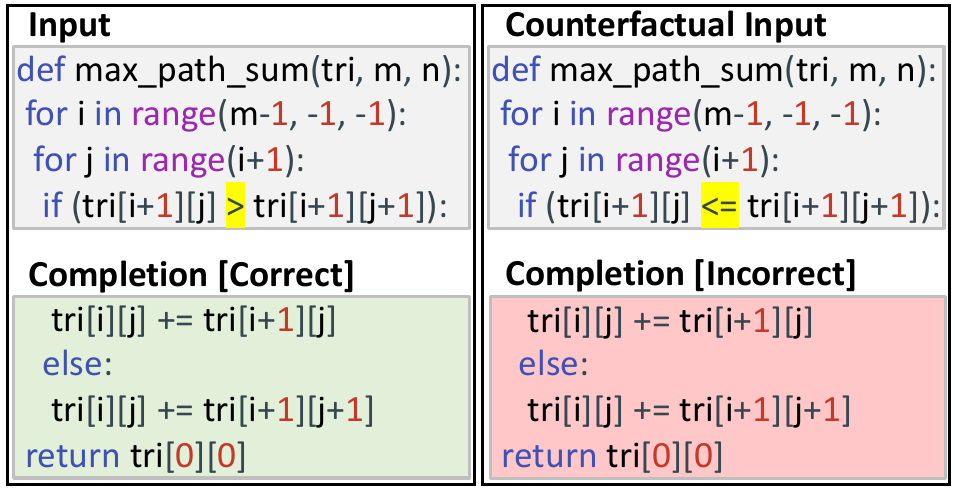}

  \caption{In this example the counterfactual input is generated by negating the relational expression in the \lstinline{if} statement. Starcoder~\cite{li2023starcoder} generates an incorrect completion for the input on the right. This suggests that LLMs have incomplete understanding of programming concepts such as \textit{control-flow}.}
  \label{fig:cfinstance}
\end{figure}

Our evaluations of ten popular LLMs reveal that state-of-art completion models have gaps in understanding \predicates, where some mutations result in more than 20\% of the tasks completed with incorrect code. \autoref{fig:cfinstance} shows an example generated by our framework, where flipping an if-condition results in an incorrect code output.

In summary, our work makes the following contributions:
\begin{enumerate}[itemsep=0.25ex,topsep=0.25ex,leftmargin=1.5em]
    \item We propose \name, a counterfactual testing framework for evaluating understanding of Programming Concept Predicates (\predicates). We show how to overcome challenges of generating counterfactual programs.
    
    \item We apply \name to the code completion task and test four types of \predicates. To this end, we extend three popular code datasets---HumanEval, MBPP, and CodeContests---and create a new benchmark dataset for evaluating \predicate understanding in LLMs.
    \item Using \name, we evaluate ten popular LLMs and provide insights on how the model's understanding depends on different model and data parameters. We highlight the gaps in the state-of-art models' understanding of coding concepts.
\end{enumerate}

\section{Background and Related Work}

\paragraph{Programming Concept Predicates and LLMs for Code.}

Programming Concept Predicates describe properties of specific elements of the program (variables, functions, data values, execution paths, etc.) either by themselves or in relation to other elements~\cite{10.1145/363235.363259}. For example, a predicate may describe the range of values a variable \lstinline{v} may take at a program location $l$, or whether some execution from location $l_1$ in function $f_1$ could reach location $l_2$ in function $f_2$ (these are a type of \textit{control-flow predicates}), or whether the value assigned to variable \lstinline{w} at location $l_1$ could be the value used when \lstinline{w} is later accessed at location $l_2$ (a type of \textit{data-flow predicate}). We say a program satisfies a predicate if in every possible execution of that program the predicate evaluated over the actual values of the relevant program elements is true\footnote{For our purposes, describing \predicates as holding over all program executions is without loss of generality, as the predicate itself may limit its scope to some subset of executions.}.

Large language models (LLMs) have shown strong performance on a variety of code tasks, from code completion~\cite{austin2021program,fried2022incoder}, to code translation~\cite{pan2023understanding,chen2023teaching}, and to code repair~\cite{jiang2021cure,joshi2023repair}. A code LLM takes as input a sequence of natural-language instructions and a sequence of code statements (i.e., a partial program) and outputs another partial program (depending on the task). We consider the general case where the task of interest has an associated function (called the \textit{attribution function}) that determines whether the output of the model satisfies the input instruction. For generative tasks for code such as code completion or code repair, it is common to use program testing as attribution function, where the output program is executed against a test suite.

The core problem we investigate is how to estimate a model's understanding of \predicates. Such an estimation can be useful to validate a model's suitability for a particular task, where the task is expected to depend (or not depend at all) on a particular predicate. For example, the task of code completion is useful only when it is sensitive to the order of program statements and thus it is expected to depend on control-flow predicates. In turn, a model trained for code completion should yield different outputs on programs with statements in different orders. If a task depends on a predicate, we want any model trained for that task to have high understanding of the predicate.

\paragraph{Robustness of Code Models.}
Recent work has studied the robustness of code models against both natural and adversarial perturbations. \citet{shirafuji2023exploring} \& \citet{wang2022recode} study robustness of code completion against different representations for the problem description as well as the input program. \citet{henkel2022semantic, jha2023codeattack} demonstrate that function name prediction models can be attacked using semantic preserving transformation applied to the input program. \citet{chen2023evaluating} have similar findings for the code summarization task. In this work, we focus on evaluating the understanding of specific programming concepts. Our approach is based on causal analysis, which involves carefully designing counterfactuals and attributing their effect. Similar to robustness, our approach also involves mutating programs and performing inference. However, our mutations are aimed at generating counterfactuals and need to ensure that the input for the original prompt and the counterfactual prompt differ only along the concept to be tested.

\paragraph{Counterfactual Analysis.}
For ML models, counterfactual analysis proceeds by performing interventions on the inputs and observing the changes in the model outputs. This can be achieved via counterfactual (CF) inputs generated by changing an input $x$ such that only a specific concept $C_k$ of the input is changed to a different value i.e. $x_{C_k=c'}$ is a counterfactual for input $x_{C_k=c}$ for concept $C$. Now, the effect of the concept on the model can be estimated by observing how the model output differs from the counterfactual. To be effective, CFs are designed to achieve three main properties~\cite{abid2022meaningfully} --- (1) \underline{Correctness}: CF perturbations should lead to a predictable change in the ground-truth output, (2) \underline{Validity}: CFs should pertain to real world constraints, and (3) \underline{Specificity}: CFs should only perturb individual properties in order to evaluate understanding of specific concepts.

In contrast to tabular and image data, generating counterfactuals has been relatively unexplored for programs. Past work on counterfactual explanations for code has looked only into syntactic perturbations and has primarily focused on finding the minimum perturbations that change the output~\cite{cito2022counterfactual}. Since these perturbations do not change isolated concepts, they are more useful in explaining model behaviour for individual inputs rather than evaluating understanding of specific concepts. In contrast, we focus on both syntactic and semantic perturbations that only change programs along specific \predicates.

Independently, there has been work on counterfactual analysis of output token probabilities of large code models~\cite{palacio2023toward,palacio2023evaluating}. These methods only work for the next predicted token and do not apply to outputs with multiple tokens. They also require access to the probability distribution of the output token prior to sampling. In contrast, our method works for the entire output and works in the hard label black box setting with access only to the final output.

\section{Counterfactual Analysis for Programming Concept Predicates}
\label{sec:cacp}

In the following, we describe \name, starting with the basic notation. Second, we discuss the requirements associated with counterfactual analysis for \predicates. Third, we describe how \name addressed these challenges for four \predicates. Finally, we describe how \name estimates the model's understanding.

\subsection{Notation}
Let $\mathsf{M}$ be a code LLM such that
$$\mathsf{M} : \mathcal{H} \times \mathcal{X} \rightarrow \mathcal{Y},$$
where $\mathcal{H}$ is the space of instructions and $\mathcal{X}, \mathcal{Y} \in \mathcal{P}$ with $\mathcal{P}$ being the space of programs. For code completion, $\mathcal{H}$ is the docstring or the problem specification in natural language, and $\mathcal{X}$ and $\mathcal{Y}$ are program prefixes and completions, respectively. An attribution function $\mathsf{A} :\mathcal{H} \times \mathcal{X} \times \mathcal{Y} \rightarrow \{0,1\}$ evaluates if the model output satisfies the instruction. 
Also, let $O_{h\times x} = \{y \;|\; y \in \mathcal{Y}, \mathsf{A}(h,x,y) = 1 \}$  be the set of correct outputs for a given instruction-input pair, where $x \in \mathcal{X}, h \in \mathcal{H}$. For code completion, a common attribution function evaluates if the completed program passes the unit tests specified by the problem.

\subsection{Requirements}

We now describe the requirements, and related challenges, for generating counterfactual programs~\cite{abid2022meaningfully}.

\begin{enumerate}[itemsep=0.25ex,topsep=0.25ex,leftmargin=1.5em]
    \item \textbf{Correctness:} A counterfactual needs to correctly solve the original task. For programs, this would mean that the perturbed program should still be able to solve the task described by the instructions. We use the task's attribution function to verify this condition. Specifically, for a model $\mathsf{M}$, a counterfactual pair $x,x' \in \mathcal{X}$, associated problem description $h \in \mathcal{H}$ and corresponding attribution function $\mathsf{A}$, we ensure that $|O_{h\times i}| > 0 \;\; \forall i \in \{x,x'\}$.
    
    \item \textbf{Validity:} The generated counterfactuals also need to be valid, i.e., they need to pertain to real-world constraints. This means that the perturbed programs should be syntactically correct. Furthermore, they should be ``natural,'' i.e., in distribution with programs seen in the software development pipeline~\cite{hindle2016naturalness}. 
    
    \item \textbf{Specificity:} Counterfactual perturbations should only change specific attributes/concepts in the input, which is especially challenging for programs. Formally, let $\mathit{Preds}(x)$ be the infinite set of all \predicates that a program $x \in \mathcal{X}$ satisfies. Note that $\mathit{Preds}(x)$ is infinite because for any predicates $\mathsf{p}_1$ and $\mathsf{p}_2$ in $\mathit{Preds}(x)$, the predicates $\mathsf{p}_1 \vee \mathsf{p}_2$ and $\mathsf{p}_1 \wedge \mathsf{p}_2$ are also in $\mathit{Preds}(x)$. This implies that any mutation applied to the program $x$ cannot affect exactly one predicate $\mathsf{p} \in \mathit{Preds}(x)$, but rather it affects a subset of $\mathit{Preds}(x)$. Therefore, for programs, we relax this requirement by considering counterfactuals that affect only a minimal set of \predicates.
\end{enumerate}

\begin{figure*}[ht]
    \centering
    \includegraphics[width=\textwidth]{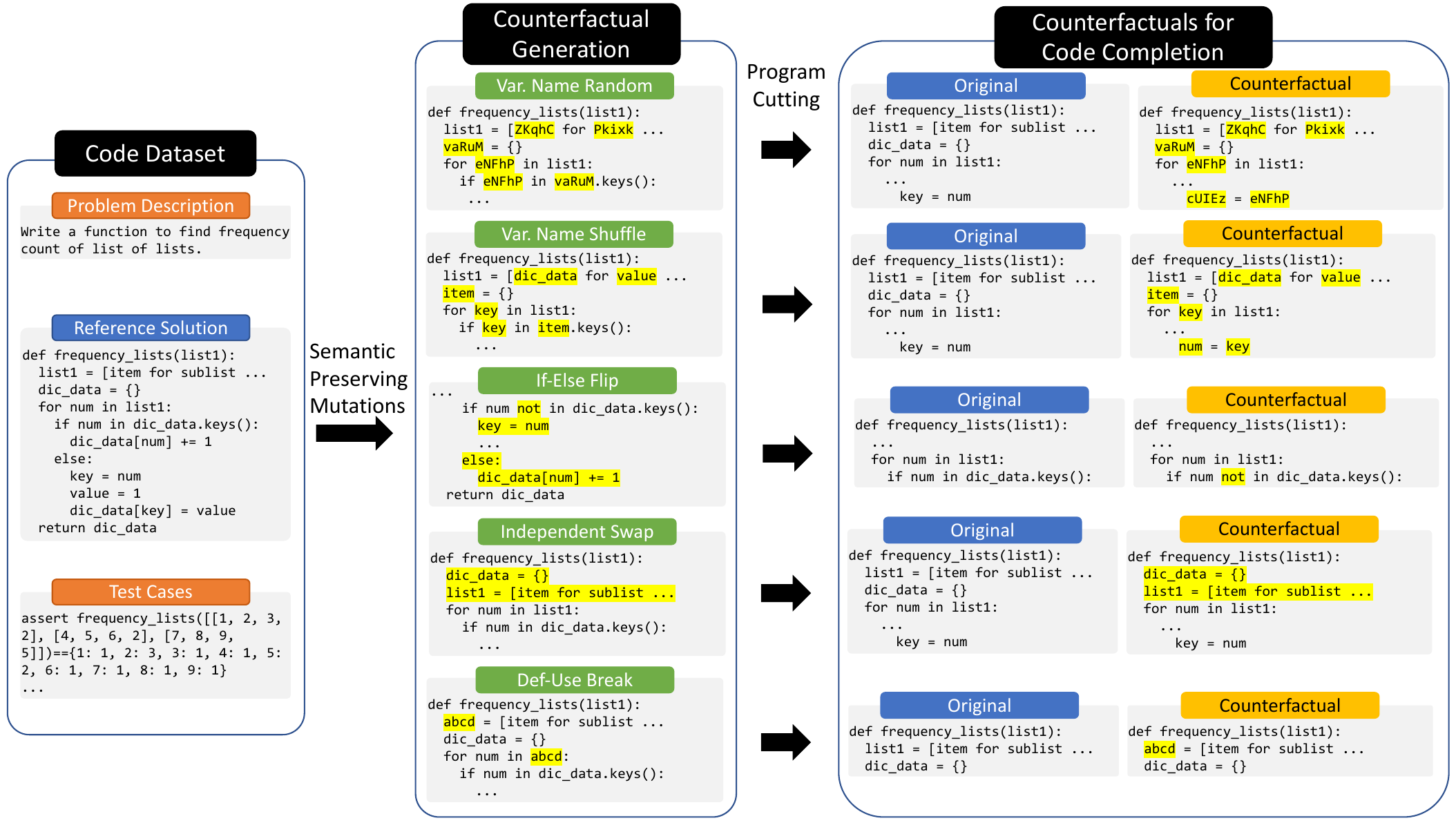}
    \caption{The counterfactual generation pipeline of \name. It consists of two stages. First, the reference solution for the problem is perturbed using predicate-specific mutations. Second, both the original and the perturbed solution are cut at the same location to generate a pair of counterfactual inputs. }
    \label{fig:cfgeneration}
\end{figure*}

\subsection{Mutations for Counterfactual Programs}
Now, we discuss how \name generates counterfactual programs that satisfy the above requirements. \name automates the CF generation process using \textit{mutations}. These are transformation functions that perturb programs with respect to specific concepts, i.e., $\mathsf{T}_{\mathsf{p}_k} : \mathcal{X} \rightarrow \mathcal{X}$ where $\mathsf{p}_k$ is the target \predicate. A \predicate can have more than one associated mutation. 
Given an input program $x \in \mathcal{X}$, the mutation function is then used to generate a counterfactual $x_{\mathsf{p}_k} = \mathsf{T}_{\mathsf{p}_k}(x) \in \mathcal{X}$. Our comprehensive review of the program analysis literature revealed four themes of studied program predicates: control flow~\cite{allen1970control,yang2015static}, data flow~\cite{fosdick1976data,nilsson2009declarative}, identifier names~\cite{lin2008evaluation}, and data types~\cite{dart1992efficient,allamanis2020typilus}. As we study weakly typed programs (for instance, Python), we consider four distinct PCPs that cover the first three themes. Next, we show how \name automates the generation of these four distinct PCPs (also illustrated in \autoref{fig:cfgeneration}).

\noindent \textbf{If-Else Flip:} We use a mutation that swaps the branches of an \lstinline{if-else} statement and negates the condition to test for the \predicate: \textit{Inverting the relational expression of an if-else block flips the ordering of the then and else bodies}. 
It involves two steps: Negating the test condition of the \lstinline{if-else} statement using DeMorgan's law and swapping the \lstinline{then} body with the \lstinline{else} body. This mutation satisfies -- (1) \underline{Correctness}: The counterfactual still solves the task since it is semantically equivalent to the input; (2) \underline{Validity}: We negate the relational expression by using complementary operators, for example, we substitute \lstinline{x==y} with \lstinline{x!=y}; (3) \underline{Specificity}: We ensure that we do not affect other \predicates by only applying this perturbation to relational expressions that do not include any method calls that might change the state of the program.

\noindent \textbf{Independent Swap:} Next, we evaluate the \predicate: \textit{Code Completion is invariant to the ordering of independent statements}. This mutation swaps pairs of independent statement blocks in the program. We use data-flow analysis to identify pairs of independent blocks. This mutation satisfies -- (1) \underline{Correctness}: Since we only swap independent blocks, the perturbed program is semantically identical and still solves the problem; (2) \underline{Validity}: Ordering of independent statements does not change the ``naturalness'' of the program; (3) \underline{Specificity}: Our data-flow analysis ensures that we only swap statements where the ordering does not affect any other \predicate.

\noindent \textbf{Def-Use Break:}
We design a mutation that breaks def-use chains to evaluate the \predicate: \textit{Breaking a def-use chain alters the scope of variables}. Def-Use chains capture the relationship between the definitions of variables (where a variable is assigned a value) and their subsequent uses (where that value is accessed or modified). To break a def-use chain, we substitute a variable's second chain with a new name (a random string of 5 characters), i.e., we simply rename the second definition and all subsequent uses. For example, in \autoref{fig:cfgeneration}, we rename the second chain of variable \lstinline{list1}. This mutation satisfies -- (1) \underline{Correctness}: we ensure that the counterfactual is semantically equivalent and still solves the problem by consistently substituting all subsequent occurrences; (2) \underline{Validity}: Random strings are often used as identifiers in obfuscated or minified versions of programs~\cite{8812034}; (3) \underline{Specificity}: We use def-use analysis to identify and perturb individual chains.

\noindent \textbf{Variable-Name Invariance:} Next, we evaluate the \predicate: \textit{Variable names do not affect the
semantics of a program}. Here, we generate
counterfactuals by renaming variables. We consider two variants of this
mutation --- renaming to random strings and permuting or shuffling existing names between variables. For the first variant, we substitute variable names with randomly generated strings of five characters. For the second variant, we shuffle names among the variables defined in the program. This mutation satisfies -- (1) \underline{Correctness}: we ensure that the counterfactual is semantically equivalent by consistently substituting each variable; (2) \underline{Validity}: We only substitute user-defined variables and do not rename reserved keywords; (3) \underline{Specificity}: We do not substitute function parameters as their names decide the order in which arguments are passed during invocation.

\subsection{Measuring Counterfactual Effect}
We need a way to analyze the effect of mutations on the observed output. For a single program $x \in \mathcal{X}$, instruction $h \in \mathcal{H}$, attribution function $\mathsf{A}$, and model $\mathsf{M}$, we formulate the mutation effect (\textsf{ME}) as: 
$$\mathsf{ME}^\mathsf{M}_{(\mathsf{p}_k,h,x)} = | \mathsf{A}(h, x_{\mathsf{p}_k}, \mathsf{M}(h,x_{\mathsf{p}_k})) - \mathsf{A}(h, x, \mathsf{M}(h,x))|$$

For code completion, a model that understands: \textit{Variable names do not affect the semantics of a program} would generate a correct completion even for the renamed program, leading to a mutation effect of 0. A model that relies on variable names might generate erroneous completions, leading to a mutation effect of 1. To compute the $\mathsf{ME}$ across all programs, we define the Average Mutation Effect ($\mathsf{AME}$):
$$\mathsf{AME}^\mathsf{M}_{\mathsf{p}_k} = \underset{h,x \in \mathcal{H},\mathcal{X}}{\mathbb{E}} \left[
\mathsf{ME}^\mathsf{M}_{(\mathsf{p}_k,h,x)}\right]$$

$\mathsf{AME}$ with a small magnitude indicates a better understanding of the \predicate. On the other hand, a large magnitude indicates poor understanding since the model performs worse after the mutation. Note that this formulation is similar to the Average Treatment Effect used in counterfactual analysis~\cite{pearl2009causal}. The treatment Effect is defined for the output of the model, whereas we compute the Mutation Effect using the attribution function.

\section[CACP for Code Completion]{\name for Code Completion}\label{sec:codecompletion}

In this section, we instantiate \name for the Code Completion task.
We first briefly
describe the code completion task. Then, we demonstrate how \name generates counterfactuals for code completion for the four \predicates. Finally, we describe how we measure the mutation effect.

\subsection{Large Language Models for Code Completion}
Code completion tasks, such as HumanEval~\citep{chen2021evaluating} and MBPP~\citep{austin2021program}, have become instrumental in evaluating the capabilities of code completion models. These tasks challenge models with an array of programming tasks designed to test different aspects of coding proficiency. In these benchmarks, problems are presented as Python \textit{function skeletons} with accompanying \textit{descriptions} that specify what the function should accomplish, along with \textit{unit tests} to validate the correctness of the generated code. Each problem in these benchmarks is also accompanied by a \textit{reference solution} that acts as a gold standard, allowing for direct comparison between model-generated code and the expected output. 

While HumanEval and MBPP excel in testing a model's ability to generate syntactically and semantically correct code, they do not assess the model's understanding of \predicates. To address this gap, \name extends these datasets by using reference solutions as a base and generating counterfactuals that can be used to evaluate the understanding of specific \predicates.

\subsection[CACP Counterfactual Generation]{\name Counterfactual Generation}\label{subsec:implementation}
\name generates counterfactuals for code completion using a two-step procedure: (1) Reference solutions are transformed using mutations specified in Section \ref{sec:cacp} to generate mutated solutions, and (2) Reference and mutated solutions are cut at the same location to create partial programs which act as counterfactual inputs. Additionally, we test these mutated solutions by compiling and executing them to confirm that they pass the required test cases. Below, we describe how we cut the solutions for each mutation (also illustrated in \autoref{fig:cfgeneration}):

\noindent \textbf{If-Else Flip:} We cut both the reference solution as well as the perturbed solution at the beginning of the \lstinline{then} body. As shown in \autoref{fig:cfgeneration}, this generates partial programs which end at a statement of the form - \lstinline{if <condition>} and the relational condition for the counterfactual is the negation of the original.

\noindent \textbf{Independent Swap:} We only consider mutations where both the swapped statements are part of the initial 75\% of the program. Then, we cut the trailing 25\%, and the remaining acts as the input for the code completion task. Note that the cutting for both the original and the counterfactual happens at the same location since the ordering of statements after the swapped pair does not change. 

\noindent \textbf{Def-Use Break:} We only consider mutations where the perturbed chain is at least partially present in the initial 75\% of the program. Then, we cut trailing 25\% for both the original and the counterfactual. This ensures that counterfactual input is not identical to the original. Note that the cutting happens at the same location since renaming the variable does not affect the line numbers of statements.

\noindent \textbf{Variable-Name Invariance:} We only consider mutations where at least one variable appearance is renamed in the initial 75\% of the program. This ensures that counterfactual input is not identical to the original. We cut off the trailing 25\% and use the rest as the counterfactuals.

\subsection[CACP Effect Measurement]{\name Effect Measurement}
\label{sec:effect}
There are two primary approaches to evaluating the generations of a code-completion task---testing and exact string matching. Exact string-matching techniques like CodeBLEU~\cite{ren2020codebleu} and chrF~\cite{popovic-2015-chrf} evaluate generations by computing the distance from the reference solution. However, such match-based metrics are unable to account for the large space of programs that are functionally equivalent to, yet syntactically distinct from, a reference solution and thus underestimate the capabilities of a model that understands programming concepts. Testing provides a more direct evaluation, where a generation is deemed correct if it passes all the unit tests for that code-completion instance. Therefore, we use \textit{unit-test correctness} as the attribution function for computing the $\mathsf{AME}$. We generate candidate solutions by querying the model on both the original input as well as the counterfactual. Then, we execute the candidate solutions against the test cases, resulting in one of two outcomes: passing all test cases or at least one failure. Note that we only consider problems where the model generates a successful completion (i.e. passing all test cases) for the original (non-perturbed) input, the perturbed input, or both. The cases where the model fails for both the original and perturbed inputs are not necessarily informative about the impact of the \predicate, and we discard them. In that case, the perturbed inputs are not considered as counterfactual.
\section{Experiments}
Using \name, we evaluate ten popular Large Language Models against five different mutations. Our evaluation answers the following questions.

\textbf{Q1: How are leading LLMs affected by counterfactual mutations?}\\
We evaluate ten popular LLMs and show that they suffer significant drops in unit test correctness for mutations on \textsf{Variable-Names}, \textsf{IfElse-Flip}, and \textsf{DefUse-Break}, leading to $\mathsf{AMEs}$ as high as $34\%$. The effect is smaller in magnitude for \textsf{Independent-Swap}. Overall, these results suggest that current models lack understanding of program predicates.

\textbf{Q2: How does the Average Mutation Effect ($\mathsf{AME}$) depend on LLM parameters?}\\
We observe that understanding of predicates seems to improve with model size. Training or fine-tuning on code-specific data also seems to improve understanding, specifically for variable name-related predicates.

\textbf{Q3: Are the errors related? What do they depend on?}\\
We analyze the correlation between pairs of mutations and show that all pairs exhibit low correlation apart from the two \textsf{Variable Names} mutations. In the case of StarCoder~\cite{li2023starcoder}, our analysis suggests a relation between $\mathsf{AME}$ for the \textsf{IfElse-Flip} mutation and the frequency of appearance of different relational operators in the model's training data.

\begin{table}
\footnotesize
\centering
\caption{Number of valid counterfactual pairs per mutation type. }
\label{tab:cfcount}
\begin{tabular}{@{}lrrr@{}}
\toprule
\textbf{Mutation} & \multicolumn{3}{c}{Counterfactual Pairs} \\ \cmidrule{2-4}
& HumanEval+MBBP & CC & Total\\
\midrule
Var. Name Random & 724 & 1000 & 1724
\\
Var. Name Shuffle & 724 & 1000 & 1724
\\
If-Else Flip & 103 & 1000 & 1103
\\
Independent Swap & 624 & 1000 & 1624
\\
Def-Use Break & 22 & 277 & 299
\\

\bottomrule
\end{tabular}
\end{table}

\begin{table*}[t]
\small
\centering
 \caption{We compute the $\mathsf{AME}$ using the Pass/Fail attribute function as described in \autoref{sec:effect}. We only consider problems where the model achieves non zero accuracy on either the original or the counterfactual setting.}\label{tab:overall}

\begin{tabular}{@{}lllccccc}
\toprule
& & & \multicolumn{5}{c}{\textbf{Average Mutation Effect (AME)}}
\\
\cmidrule{4-8}\\
\multirow{2}{*}{\textbf{\begin{tabular}[c]{@{}c@{}}\\\ Dataset\end{tabular}}} & \multirow{2}{*}{\textbf{\begin{tabular}[c]{@{}c@{}}\\Model\end{tabular}}} & \multirow{2}{*}{\textbf{\begin{tabular}[c]{@{}c@{}} Original \\ Accuracy \end{tabular}}} &
\multirow{2}{*}{\textbf{\begin{tabular}[c]{@{}c@{}} \textsf{Variable-Names} \\ Random \end{tabular}}} &
\multirow{2}{*}{\textbf{\begin{tabular}[c]{@{}c@{}} \textsf{Variable-Names} \\ Shuffle\end{tabular}}} &
\multirow{2}{*}{\textbf{\begin{tabular}[c]{@{}c@{}} \textsf{IfElse-} \\ \textsf{Flip}\end{tabular}}} &
\multirow{2}{*}{\textbf{\begin{tabular}[c]{@{}c@{}} \textsf{Independent-} \\ \textsf{Swap} \end{tabular}}} &
\multirow{2}{*}{\textbf{\begin{tabular}[c]{@{}c@{}} \textsf{DefUse-} \\ \textsf{Break} \end{tabular}}} \\
& & & & & & \\\midrule
\multicolumn{1}{@{\ }l|}{\multirow{8}{*}{\textbf{\begin{tabular}[c]{@{}c@{}} HumanEval \\ + \\ MBPP\end{tabular}}}}  & 
Starcoder (13B) & 66.04 \% & \intensitycolor{-1}{16.86}{17} & \intensitycolor{-1}{19.42}{19} & \intensitycolor{-1}{21.07}{21} & \intensitycolor{-1}{07.47}{7} & \intensitycolor{-1}{05.00}{5} \\ 
\multicolumn{1}{l|}{} & Llama 2 (7B) & 43.20 \% & \intensitycolor{-1}{24.58}{25} & \intensitycolor{-1}{29.08}{29} & \intensitycolor{-1}{25.18}{25} & \intensitycolor{-1}{13.45}{13} & \intensitycolor{-1}{21.88}{22} \\
\multicolumn{1}{l|}{} & Llama 2 (13B) & 48.40 \% & \intensitycolor{-1}{21.14}{21} & \intensitycolor{-1}{26.84}{27} & \intensitycolor{-1}{20.00}{20} & \intensitycolor{-1}{09.12}{9} & \intensitycolor{-1}{15.88}{16} \\
\multicolumn{1}{l|}{} & Llama 2 (70B) & 63.37 \% & \intensitycolor{-1}{14.37}{14} & \intensitycolor{-1}{19.81}{20} & \intensitycolor{-1}{20.83}{21} & \intensitycolor{-1}{05.54}{6} & \intensitycolor{-1}{06.50}{7} \\
\multicolumn{1}{l|}{} & Llama Code (7B) & 60.10 \% & \intensitycolor{-1}{19.84}{20} & \intensitycolor{-1}{21.44}{21} & \intensitycolor{-1}{17.71}{18} & \intensitycolor{-1}{10.88}{11} & \intensitycolor{-1}{05.00}{5} \\
\multicolumn{1}{l|}{} & Llama Code (13B) & 66.61 \% & \intensitycolor{-1}{12.56}{13} & \intensitycolor{-1}{18.06}{18} & \intensitycolor{-1}{16.62}{17} & \intensitycolor{-1}{05.04}{5} & \intensitycolor{-1}{09.50}{10} \\
\multicolumn{1}{l|}{} & Llama Code (34B) & 72.65 \% & \intensitycolor{-1}{12.55}{13} & \intensitycolor{-1}{15.14}{15} & \intensitycolor{-1}{17.09}{17} & \intensitycolor{-1}{04.76}{5} & \intensitycolor{-1}{07.62}{8} \\
\multicolumn{1}{l|}{} & PaLM 2 (64B) & 45.74 \% & \intensitycolor{-1}{23.75}{24} & \intensitycolor{-1}{22.58}{23} & \intensitycolor{-1}{25.00}{25} & \intensitycolor{-1}{12.96}{13} & \intensitycolor{-1}{19.38}{19} \\
\multicolumn{1}{l|}{} & PaLM 2 (340B) & 66.98 \% & \intensitycolor{-1}{14.71}{15} & \intensitycolor{-1}{17.70}{18} & \intensitycolor{-1}{19.72}{20} & \intensitycolor{-1}{06.13}{6} & \intensitycolor{-1}{17.00}{17} \\
\multicolumn{1}{l|}{} & PaLM 2-$S^*$ (24B) & 70.01 \% & \intensitycolor{-1}{12.31}{12} & \intensitycolor{-1}{19.74}{20} & \intensitycolor{-1}{16.09}{16} & \intensitycolor{-1}{06.51}{6} & \intensitycolor{-1}{11.90}{12} \\
\multicolumn{1}{l|}{} & GPT4 (gpt-4-1106) & 88.94 \% & \intensitycolor{-1}{06.43}{6} & \intensitycolor{-1}{07.21}{7} & \intensitycolor{-1}{05.95}{6} & \intensitycolor{-1}{01.57}{2} & \intensitycolor{-1}{04.76}{5} \\

\midrule
\multicolumn{1}{@{\ }l|}{\multirow{8}{*}{\textbf{\begin{tabular}[c]{@{}l@{}} Code \\ \multicolumn{1}{@{}r@{}}{\quad Contests} \end{tabular}}}}  & 
Starcoder (13B)& 43.75 \% & \intensitycolor{-1}{16.90}{17} & \intensitycolor{-1}{21.18}{21} & \intensitycolor{-1}{30.93}{31} & \intensitycolor{-1}{06.43}{6} & \intensitycolor{-1}{22.92}{23} \\ 
\multicolumn{1}{l|}{} & Llama 2 (7B) & 24.75 \% & \intensitycolor{-1}{29.14}{29} & \intensitycolor{-1}{25.38}{25} & \intensitycolor{-1}{29.72}{30} & \intensitycolor{-1}{13.24}{13} & \intensitycolor{-1}{34.07}{34} \\
\multicolumn{1}{l|}{} & Llama 2 (13B) & 29.48 \% & \intensitycolor{-1}{23.78}{24} & \intensitycolor{-1}{23.86}{24} & \intensitycolor{-1}{29.52}{30} & \intensitycolor{-1}{09.26}{9} & \intensitycolor{-1}{23.98}{24} \\
\multicolumn{1}{l|}{} & Llama 2 (70B) & 40.18 \% & \intensitycolor{-1}{17.19}{17} & \intensitycolor{-1}{18.20}{18} & \intensitycolor{-1}{28.58}{29} & \intensitycolor{-1}{09.14}{9} & \intensitycolor{-1}{26.04}{26} \\
\multicolumn{1}{l|}{} & Llama Code (7B) & 38.74 \% & \intensitycolor{-1}{22.16}{22} & \intensitycolor{-1}{21.62}{22} & \intensitycolor{-1}{26.95}{27} & \intensitycolor{-1}{09.21}{9} & \intensitycolor{-1}{20.23}{20} \\
\multicolumn{1}{l|}{} & Llama Code (13B) & 40.66 \% & \intensitycolor{-1}{21.45}{21} & \intensitycolor{-1}{22.52}{23} & \intensitycolor{-1}{32.53}{33} & \intensitycolor{-1}{07.48}{7} & \intensitycolor{-1}{29.40}{29} \\
\multicolumn{1}{l|}{} & Llama Code (34B) & 49.55 \% & \intensitycolor{-1}{16.53}{17} & \intensitycolor{-1}{18.09}{18} & \intensitycolor{-1}{32.02}{32} & \intensitycolor{-1}{07.04}{7} & \intensitycolor{-1}{26.60}{27} \\
\multicolumn{1}{l|}{} & PaLM 2 (64B) & 38.75 \% & \intensitycolor{-1}{18.18}{18} & \intensitycolor{-1}{21.53}{22} & \intensitycolor{-1}{26.43}{26} & \intensitycolor{-1}{08.06}{8} & \intensitycolor{-1}{23.11}{23} \\
\multicolumn{1}{l|}{} & PaLM 2 (340B) & 47.27 \% & \intensitycolor{-1}{15.57}{16} & \intensitycolor{-1}{17.90}{18} & \intensitycolor{-1}{27.31}{27} & \intensitycolor{-1}{07.58}{8} & \intensitycolor{-1}{18.56}{19} \\
\multicolumn{1}{l|}{} & PaLM 2-$S^*$ (24B) & 47.28 \% & \intensitycolor{-1}{13.22}{13} & \intensitycolor{-1}{15.59}{16} & \intensitycolor{-1}{29.37}{29} & \intensitycolor{-1}{05.48}{5} & \intensitycolor{-1}{18.25}{18}
                                                   \\
\multicolumn{1}{l|}{} & GPT4 (gpt-4-1106) & 67.83 \% & \intensitycolor{-1}{11.25}{11} & \intensitycolor{-1}{16.48}{16} & \intensitycolor{-1}{14.89}{15} & \intensitycolor{-1}{05.05}{5} & \intensitycolor{-1}{21.58}{22} \\
\bottomrule
\end{tabular}
\end{table*}

\subsection{Experimental Setup}
We use the following settings to demonstrate how \name evaluates understanding of programming concepts. 

\noindent \textbf{Datasets and mutations.} We instantiate \name using three popular code generation benchmarks --- HumanEval~\cite{chen2021evaluating}, MBPP~\cite{austin2021program}, and CodeContests~\cite{doi:10.1126/science.abq1158}. All of the problems in these datasets include a reference solution, which is used to generate counterfactual pairs as described in \autoref{sec:codecompletion}. Since not every mutation applies to all reference solutions, the final number of counterfactual pairs differs based on the mutation type. As shown in \autoref{tab:cfcount}, mutations related to \textsf{Variable Names} can be applied to almost all solutions, whereas mutations related to control-flow or def-use are more selective. In this evaluation, we focus on Python programs, but our methodology applies to any programming language. We use \texttt{libCST} \cite{libcst_docs} for parsing and manipulating source code for our mutations.

\noindent \textbf{Models.}
We use \name to evaluate popular models, including Llama 2~\cite{touvron2023llama} and PaLM~\cite{anil2023palm}. We also evaluate counterparts of these models that are fine-tuned for coding tasks -- Code Llama~\citep{roziere2023code} and PaLM 2-$S^*$~\citep{anil2023palm}. Finally, we also evaluate the popular open source code LLM StarCoder~\citep{li2023starcoder}. 
We set the sampling temperature to 0 for all models to have deterministic results. 

\begin{figure*}[t]
  \centering

  \includegraphics[width=\textwidth]{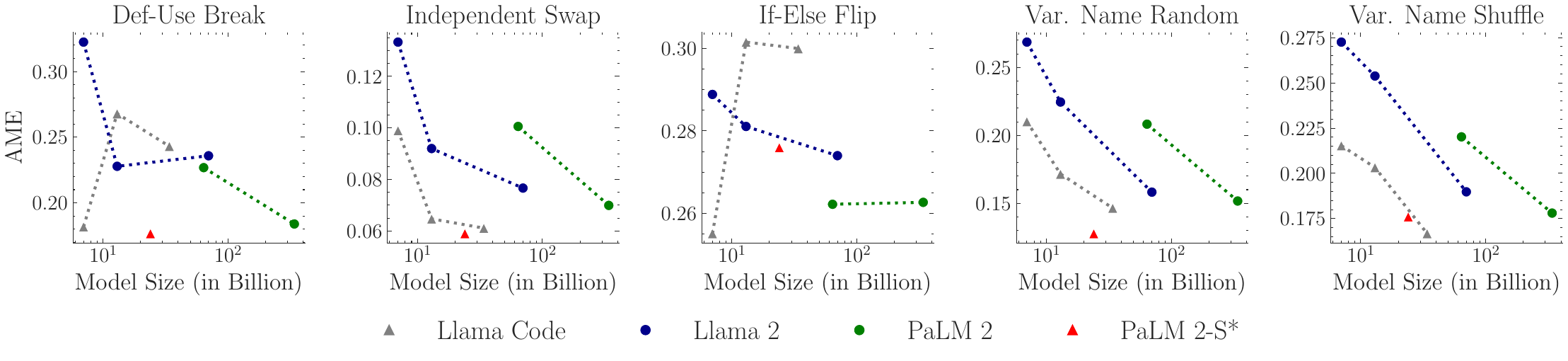}

  \caption{$\mathsf{AME}$ as a function of model size (number of parameters in Billions). The different model classes are depicted using different colors.}
  \label{fig:scale}
\end{figure*}

\begin{figure}[t]
  \centering

  \includegraphics[width=0.5\columnwidth]{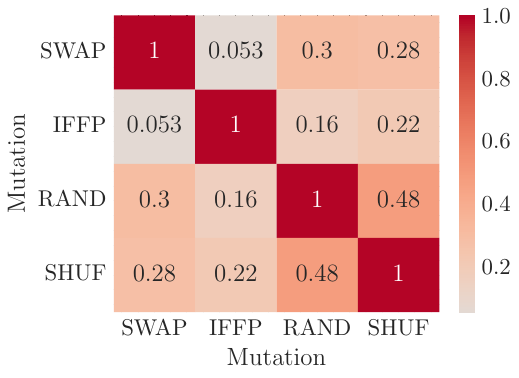}
    
  \caption{Correlation between $\mathsf{AME}$ values across pairs of mutations. The number of samples used to compute each value depends on the size of the intersection of the two mutation types. \textsf{Independent-Swap}: SWAP, \textsf{IfElse-Flip}: IFFP, \textsf{Variable Names Random}: RAND, \textsf{Variable Names Shuffle}: SHUF}
  \label{fig:corr}
\end{figure}

\subsection{Average Mutation Effect}

\autoref{tab:overall} shows the $\mathsf{AME}$ for the three datasets, five mutations, and ten models. The table shows that the original unit test correctness rates vary across models. $\mathsf{AME}$ values are non-zero, which suggests that models do not fully understand the evaluated \predicates. 
In the case of the \textsf{Variable-Names} and \textsf{IfElse-Flip} perturbations, $\mathsf{AME}$ values are as high as $33\%$. On the other hand, the \textsf{Independent-Swap} mutation is the most well-understood. While most mutations have similar effects across the two kinds of datasets, the \textsf{DefUse-Break} perturbation shows a relatively lower effect on the HumanEval and MBPP datasets. This is likely due to the small number of valid problems --- only 22.

\noindent \textbf{Across Models:} 
For \textsf{Variable-Name} related perturbations, we first observe that smaller models perform worse and larger models do better. This is evident in \autoref{fig:scale}, which shows the $\mathsf{AME}$ as a function of the model size. Secondly, models trained on code (StarCoder) or fine-tuned on code (Llama Code, PaLM 2-$S^*$) perform better than models that are not. Perturbations related to control flow and data flow follow a similar trend for model size, but code fine-tuning does not always seem to improve performance. GPT4 performs much better on HumanEval and MBPP, but is similar to the other models for the CodeContests dataset.

\noindent \textbf{Correlation across Mutations:} Until now, we have seen the average effect of the perturbations across the datasets. \autoref{fig:corr} shows the correlation between different perturbation types. As expected, the two \textsf{Variable-Names} perturbations correlate highly. Other perturbations have fairly low correlation, suggesting that our mutations are predicate-specific and have minimal correlated errors. 

\noindent \textbf{Errors due to Memorization:}
We performed an additional experiment to gain some insights on whether 
memorization~\cite{carlini2022quantifying} contributes to the observed mutation effects. For the \textsf{If-Else} perturbation, we analyze the connection between the frequency of appearance of relational operators in the training set and their respective change in unit test correctness. We perform this analysis with StarCoder's training data~\cite{husain2019codesearchnet}. More specifically, in \autoref{tab:memorization}, we show the relative frequency of complement relational operators and the change in correctness values when substituted. We can see that operators that appear more frequently in the training set face a significantly higher drop in correctness when they are being substituted.
\begin{table}[t]
\small
\centering
\caption{Memorization Analysis for the \lstinline{If-Else} mutation for Starcoder. We parse Starcoder's training data and show the relative frequency of appearance of pairs of complementary relational operators. We also show the average change in unit test correctness computed over all valid programs in HumanEval, MBPP and CodeContests.}
\label{tab:memorization}
\begin{tabular}{@{}cccccc@{}}
\toprule

\textbf{Op A}                      & \textbf{Op B} & \textbf{Ratio} & \textbf{$\Delta$(A$\rightarrow$B)} & \textbf{$\Delta$(B$\rightarrow$A)}\\
\midrule
$==$ & $!=$ & 3.9 & \intensitycolor{-1}{13.21}{13} & \intensitycolor{-1}{07.37}{7} \\
$>$ & $<=$ & 3.8 & \intensitycolor{-1}{16.92}{17} & \intensitycolor{-1}{01.48}{1} \\
$<$ & $>=$ & 2.2 & \intensitycolor{-1}{05.00}{5} & \intensitycolor{-1}{0.00}{0} \\
\bottomrule
\end{tabular}
\end{table}

\noindent \textbf{Effect of cutoff point:}
In \autoref{subsec:implementation}, we describe that we keep 75\% of the program as the prefix for generating counterfactuals for the \textsf{Independent-Swap}, \textsf{DefUse-Break} and \textsf{Variable Name Invariance} mutations. To study the effect of the cutoff point, we evaluated counterfactuals generated using the same set of programs but cut at different places. In \autoref{tab:cutoff}, we present the original accuracy and the $\mathsf{AME}$ for the Starcoder model. This does not include the \textsf{IfElse-Flip} mutation since in that case the cut depends on the location of the \textsf{If} block. We find that an earlier cut leads to a decrease in the original accuracy as well as a higher $\mathsf{AME}$. This is expected since cutting earlier increases the complexity of the completion task. However, we observe that $\mathsf{AME}$ is relatively more stable than the original accuracy. This suggests that $\mathsf{AME}$ is a good measure of the model's understanding, irrespective of the complexity of the coding task.

\noindent \textbf{Code Repair Task:}
We also evaluate \name for the code repair task. We use the HumanEvalPack \cite{muennighoff2023octopack} dataset which is an extension of HumanEval to also include the Code Repair task. This dataset is constructed by manually adding a bug to each solution in HumanEval. For this task, the model is tasked with fixing the bug and generating the correct solution. In this case, we generate counterfactuals by applying mutations on the buggy solution. In \autoref{tab:code_repair}, we show the performance of Octocoder \cite{muennighoff2023octopack} which is an instruction-tuned version of Starcoder. Similar to code completion, we observe a high average mutation effect which suggests a lack of understanding for the Code Repair task as well.

\section{Future Work}
\noindent \textbf{Automating Semantic Preserving Perturbations.}
 Currently, crafting these perturbations requires a significant amount of manual effort and deep domain knowledge to ensure they do not alter the underlying logic of the program and only change specific predicates. Developing automated tools and techniques that can reliably generate such perturbations will not only streamline the evaluation process but also enhance the scalability of our testing framework.
 
 \begin{table}[t]
\small
\centering
\caption{$\mathsf{AME}$ for different cutoff settings when evaluating Starcoder. A lower prefix ratio implies an earlier cut. \textsf{Independent-Swap}: SWAP, \textsf{Variable Names Random}: RAND, \textsf{Variable Names Shuffle}: SHUF, \textsf{DefUse-Break}: DUBR}
\label{tab:cutoff}
\begin{tabular}{@{}ccccccc@{}}
\toprule

\textbf{\begin{tabular}[c]{@{}c@{}} Prefix \\ Ratio \end{tabular} }                      & \textbf{\begin{tabular}[c]{@{}c@{}} Orig. \\ Acc. \end{tabular}} & \textbf{RAND} & \textbf{SHUF} & \textbf{SWAP} & \textbf{DUBR}\\
\midrule
0.4 - 0.6 & 32.8 \% & \intensitycolor{-1}{18.7}{19} & \intensitycolor{-1}{21.4}{21} & \intensitycolor{-1}{08.0}{8} & \intensitycolor{-1}{26.0}{26} \\
0.6 - 0.8 & 50.1 \% & \intensitycolor{-1}{16.7}{17} & \intensitycolor{-1}{19.6}{20} & \intensitycolor{-1}{25.0}{25} & \intensitycolor{-1}{06.0}{6} \\
0.8 - 1.0 & 60.1 \% & \intensitycolor{-1}{12.4}{12} & \intensitycolor{-1}{17.7}{18} & \intensitycolor{-1}{19.6}{20} & \intensitycolor{-1}{04.0}{4} \\
\bottomrule
\end{tabular}
\end{table}

\noindent \textbf{Perturbation-based Data Augmentation.}
A promising area of future work is the application of perturbations to data augmentation to reduce the mutation effect observed in models. 
By systematically introducing perturbed data during the training phase, models could potentially develop a more nuanced understanding of code, reducing their susceptibility to errors. This approach requires careful consideration to balance the augmentation process without introducing bias or overly diluting the training data.

\begin{table}[t]
\small
\centering
\caption{$\mathsf{AME}$ for the code repair task. We evaluate OctoCoder \cite{muennighoff2023octopack} on countefactuals generated on the code repair benchmark from HumanEvalPack. \textsf{Independent-Swap}: SWAP, \textsf{Variable Names Shuffle}: SHUF, \textsf{IfElse-Flip}: IFFP}
\label{tab:code_repair}
\begin{tabular}{@{}ccccc@{}}
\toprule
 \textbf{Original Accuracy} & \textbf{SHUF} & \textbf{IFFP} & \textbf{SWAP}\\
\midrule
15 \% & \intensitycolor{-1}{36}{36} & \intensitycolor{-1}{33}{33} & \intensitycolor{-1}{15}{15} \\
\bottomrule
\end{tabular}
\end{table}

\noindent \textbf{Expanding Counterfactual Analysis with Diverse Code Datasets.}
Our framework would benefit from adding more code datasets including ones that may not support test-based attribution functions~\cite{DBLP:journals/corr/abs-2102-04664,husain2019codesearchnet}. 
This would also help increase the number of input samples for more selective perturbations like def-use chains. However, in absence of test cases, this would require the development of specialized attribution functions. Moreover, careful attention must be paid to the provenance of the data to avoid contamination of the evaluation set with examples that may have been part of the model's training set. 

\section{Conclusion}
In conclusion, we explore whether Large Code Models understand programs and propose \name, a counterfactual testing framework for evaluating understanding of program predicates. \name builds upon existing code datasets and requires only hard-label, black-box access to the model. We use \name to evaluate ten popular large code models and demonstrate that current models suffer from accuracy drops up to $33\%$ due to lack of understanding of program predicates related to control-flow and data-flow.




\section*{Acknowledgements}
We thank Saswat Anand, Sajjad Arshad and Fengguo Wei for their help with evaluating Google models. We also thank John Cyphert, Jordan Henkel, Zi Wang and Neal Mangaokar as well as the anonymous reviewers for their valuable feedback.


\section*{Impact Statement}

This paper presents work whose goal is to advance the field of Machine Learning. There are many potential societal consequences of our work, none which we feel must be specifically highlighted here


\bibliography{example_paper}
\bibliographystyle{icml2024}




\end{document}